\documentstyle[preprint,aps,epsf,floats]{revtex}
  \def \clo      { {\rm ClO}_2  }
  \def \clom     { {{\rm ClO}_2}^-}
  \def \ma       { {\rm MA}     }
  \def \ii       { {\rm I}_2    }
  \def \im       { {\rm I}^-    }
  \def \hp       { {\rm H}^+    }
  \def \siiim    { {{\rm SI}_3}^-}
  
  \def \CLO      { [{\rm ClO}_2] }
  \def \CLOM     { [{{\rm ClO}_2}^-] }
  \def \MA       { [{\rm MA}] }
  \def \II       { [{\rm I}_2] }
  \def \IM       { [{\rm I}^-] }
  \def \HP       { [{\rm H}^+] }
  \def \S        { [{\rm S}] }
  \def \SIIIM    { [{{\rm SI}_3}^-]} 
   
\begin{document}
\draft 
\tighten
\preprint{HEP/123-qed}
\title{Numerical Bifurcation Diagram for the Two-Dimensional 
Boundary-fed CDIMA System}
\author{S. Setayeshgar\footnote{Corresponding author, Present address:
Physics Department, Northeastern University, Email: 
simas@masto.physics.neu.edu,
Fax: 617 373 2943, Telephone: 617 373 2944} and M. C. Cross}
\address{
Condensed Matter Physics 114-36, California Institute of Technology,
Pasadena, CA 91125
}
\date{\today}
\maketitle
\begin{abstract}
We present numerical solution of the chlorine dioxide-iodine-malonic
acid reaction-diffusion system in two dimensions in a boundary-fed system
using a realistic model.  The bifurcation diagram for the transition from
non-symmetry breaking structures 
along boundary feed gradients to 
transverse symmetry-breaking patterns in a single layer 
is numerically determined.  
We find this transition to be discontinuous. We make connection
with earlier results and discuss prospects for
future work.   
\end{abstract}

\narrowtext


\section{Introduction}
\label{sec:intro}

In 1952, Alan Turing showed \cite{ref:Tur_52} that under certain 
conditions, reaction and diffusion processes alone could lead to
the symmetry-breaking instability of a 
system from a homogeneous state to a stationary patterned state.
The Turing instability is characterized by an intrinsic wavelength, 
in contrast to hydrodynamic instabilities, such as 
Rayleigh-B\'enard convection where the wavelength depends on cell height.
For this reason, this instability mechanism 
has particular relevance to pattern formation 
in biological systems \cite{ref:Meinhardt,ref:Murray}.  
Given the difficulties of noninvasive experiments on biological systems, 
experimental studies of Turing pattern formation have been carried 
out primarily on chemical systems.  
Even so, Turing patterns have been generated in the laboratory 
only recently, specifically in the chlorite-iodide-malonic 
acid (CIMA) chemical reaction-diffusion system 
\cite{ref:CDBD_90,ref:DCDB_91,ref:OuSw_Nat91,ref:OuSw_95}.  

Although Turing patterns have been extensively studied theoretically 
and numerically in the context of abstract models, the possibility
of comparison with controlled and reproducible experiments
provides motivation for quantitative analyses 
based on realistic models of these systems. 
Lengyel, Rabai and Epstein (henceforth referred to as LRE) 
have proposed a realistic model of the simpler 
chlorine dioxide-iodine-malonic acid (CDIMA) 
reaction-diffusion system \cite{ref:LRE_90,ref:LRE2_90}.  
The chemistry of the CDIMA and CIMA systems are related, 
and the two are similar in terms of their stationary pattern 
forming and dynamical behavior.  The potential for both 
experimental and theoretical work makes the CDIMA system well-suited 
for the study of nonequilibrium pattern formation in general.

In practice, this has not been fully realized, and 
unlike in fluid systems \cite{ref:CH_93}, experimental and theoretical efforts
in chemical systems have not been closely coupled.
First, numerical solution of 
reaction-diffusion equations using
realistic chemical parameters is computationally demanding.
In addition, the algebraic complexity of the realistic nonlinear
reaction terms renders these models unsuitable for analysis by standard
analytical tools.
Furthermore, unlike the case considered originally 
by Turing and subsequently by
others, the experimental system is not uniform.  A continuous
supply of reactants is fed into the reactor from two separately
nonreactive reservoirs to keep the system far from
equilibrium, setting up gradients in the concentrations of 
these reservoir species along the width of the reactor. 
In an earlier work \cite{ref:ssmcc1}, we used the realistic LRE model to investigate
the formation of one-dimensional stationary structures along the boundary feed
gradients and their linear instability to transverse symmetry-breaking (Turing)
patterns. Here, we extend these results by numerically solving the 
fully nonlinear reaction-diffusion equations in two dimensions.

Kadar {\sl et al.} \cite{ref:KLE_95} have also
numerically investigated two-dimensional Turing patterns 
within the LRE model, however in a closed system 
(\textsl{i.e.}, in the absence of gradients) where the 
patterns are transient by nature. 
Two-dimensional numerical simulations of Turing patterns
in ramped systems have been performed using popular abstract models, 
such as the Schnakenberg model \cite{ref:DB1_92,ref:DB2_92}, 
and the Brusselator model \cite{ref:BDD_92,ref:DBDR_95,ref:BDDW_95}, 
as well as 
the generalized Swift-Hohenberg equation \cite{ref:HMBD_95}.
These models, which are easier to implement, produce patterns 
which possess similar features to those observed in experimental systems.
However, they do not allow
quantitative comparison or prediction of new features at specific parameter
values of a real chemical system. 

In this article, we present the first numerical solution
of the LRE model in a boundary-fed system in two dimensions, 
corresponding to sustained patterns in 
thin-strip open reactors. 
We determine the branch of the bifurcation diagram corresponding to the 
transition to stripes in this
system, a result which can be directly investigated in experiments
based on the CDIMA system. In Sec.\ \ref{sec:chemmod}, 
we present the model and parameters used.  
In Sec.\ \ref{sec:nummeth}, we describe our numerical method.
We present our results in Sec.\ \ref{sec:result}, and give
conclusions and prospects for future work in Sec.\ \ref{sec:conclude}.


\section{Chemical Model}
\label{sec:chemmod}

The LRE model of the CDIMA reaction-diffusion 
system has been discussed in detail elsewhere 
\cite{ref:KLE_95,ref:ssmcc1,ref:LE_95}.
The resulting governing partial differential
equations are given below:
\begin{eqnarray}
\label{eq:full_1}
\frac{\partial \MA}{\partial t}   &=& - r_1 + D_{\ma}\nabla^2\MA,         \\
\label{eq:full_2}
\frac{\partial \II}{\partial t}   &=& - r_1 + \frac{1}{2} r_2 + 2 r_3 - r_4
                                            + D_{\ii}\nabla^2\II,          \\
\label{eq:full_3}
\frac{\partial \CLO}{\partial t}  &=& - r_2 + D_{\clo}\nabla^2\CLO,         \\
\label{eq:full_4}
\frac{\partial \IM}{\partial t}   &=&   r_1 - r_2 - 4 r_3 - r_4
                                            + D_{\im}\nabla^2\IM,     \\
\label{eq:full_5}
\frac{\partial \CLOM}{\partial t} &=& r_2 - r_3
                                            + D_{\clom}\nabla^2\CLOM,       \\
\label{eq:full_6}
\frac{\partial {\SIIIM}}{\partial t} &=& r_4.                 \\
\end{eqnarray} 
The nonlinear reaction terms are given by:
\begin{eqnarray}
	r_1 & = & \frac{k_{1a}\MA\II}{k_{1b} + \II}, \\
	r_2 & = & k_2    \CLO \IM, \\
	r_3 & = &  k_{3a} \CLOM \IM \HP+  
                          \frac{k_{3b} \CLOM \II \IM}{h+ \IM^2}, \\
	r_4 & = & k_+ \S \II \IM - k_- \SIIIM.
\end{eqnarray}
The rate and diffusion constants
used in the numerical calculations here are taken from 
Refs.\ \cite{ref:KLE_95,ref:LE_95,ref:LE_91} and are given 
in Table \ref{table:T1}. 
The left/right reservoir species are malonic acid
($\ma$)/chlorine dioxide ($\clo$)
and iodine ($\ii$), respectively.
As these species diffuse through and react in the gel reactor, 
the dynamical iodide ($\im$) and chlorite ($\clom$) species are produced.  
The starch triiodide
complex $\siiim$ is the experimentally observed species. 

The experimental CIMA reaction is similar to 
the CDIMA reaction in terms of dynamics and stationary pattern formation
\cite{ref:LE_95,ref:LKE_92,ref:LKE_93}.
However, a quantitative model of the CIMA reaction does not exist.
In particular, the role of chlorine-dioxide in this reaction is
not well understood \cite{ref:LE_95}.  
Furthermore, it has been pointed out that
in the experimental CIMA system, use of chlorite and iodide 
along with acid as reservoir species could lead to reactive 
reservoirs \cite{ref:LE_95}.  
In the CDIMA system, chlorine dioxide is non-reactive with both iodine
and malonic acid, allowing for well controlled boundary concentrations
of these species.  It is known from batch experiments and simulations
of the CDIMA system that concentrations of chlorine dioxide, iodine
and malonic acid vary little on the scale of variations in
the chlorite and iodide concentrations \cite{ref:LRE_90}.  This observation
has formed the basis of the adiabatic elimination of the former
reactants, resulting in a two-variable reduction of the full model
\cite{ref:LRE_90,ref:LE_92}, and making their 
identification as the background species
a natural one.  Hence, in the interest of aligning experiment
with numerical and theoretical work in this area, it appears
reasonable for experiments to implement the CDIMA system.


\section{Numerical Method}
\label{sec:nummeth}

A pseudospectral method was used to solve the governing partial
differential equation in two dimensions.
The physical boundary conditions are
no-flux in the $x$-direction (transverse to the gradients), and
fixed point boundary conditions in the $z$-direction (along the gradients).
In our numerical implementation, the governing equations are
cosine Fourier transformed in the $x$-direction, and each spectral mode 
is evolved in time as a one-dimensional problem in the $z$-direction.
We used a five-point finite-difference approximation
with a variable-width spatial mesh to allow better resolution of
the sharp front patterns along the gradients.  The time-stepping
scheme was Crank-Nicolson for the linear terms (implicit) 
and Adams-Bashford for the nonlinear terms (explicit), 
both second order accurate.  After each time step, the
updated solutions were transformed back into real space to
reconstruct the nonlinear terms. 

As Table\ \ref{table:T1} indicates, the large order-of-magnitude
variations in the real
chemical parameters of the LRE model make numerical solution of
this model less tractable than the aforementioned abstract models.
The stability restriction on the time step 
$\Delta t$ due to explict treatment of the nonlinear
terms is onerous.  Hence, we parallelized the above numerical
scheme, and the computations were performed on a $512$-node Intel Paragon.

As initial conditions, we used the stationary solution 
$\overrightarrow{u_s}(z)$ along $z$ 
with a uniform profile in $x$,
to which we added a perturbation given by 
the most linearly unstable mode, $k^*$:
\begin{equation}
\overrightarrow{u}(x,z,t=0)=\overrightarrow{u}_s(z)+
C\overrightarrow{\delta u}_{k^*}(z)\cos{(k^*x)}. \nonumber
\end{equation}
$\overrightarrow{\delta u}_{k^*}(z)$ is the 
eigenvector obtained from the linear
stability analysis, and
$C$ is an overall scale factor to ensure that the perturbation
is small and lies in the linear regime. 
The concentration vectors correspond to the six variables
of the LRE model.  The full six-variable linear stability 
analysis was carried out using inverse iteration \cite{ref:ssth}. 
The non-zero boundary conditions in the $z$-direction are:
$\MA_{\small \rm L}=0.0115$ M, $\CLO_{\small \rm L}=0.006$ M, 
and $\II_{\small \rm L}=0.008$ M, where $(R,L)$ refer the right and
left reservoirs, respectively.
From the linear stability analysis of $\overrightarrow{u_s}(z)$, 
the growth rate for the
instability at $k^*=471.2$ ${\rm cm^{-1}}$ is 
$\lambda(k^*)=0.00465$ ${\rm s^{-1}}$,
giving a characteristic saturation time of 
$\tau \sim 1/\lambda \sim 215$ s. 
The system size is $0.3$ cm in the $z$-direction and $0.133$ cm in
the $x$-direction, corresponding to exactly ten wavelengths in
the $x$-direction.  The spatial resolution of the system
investigated here was $N_x=129$ and $N_z=914$. 
One-thousand iterations
at this resolution required approximately $12$ node-hours.

The integration time step was $\Delta t=0.001$ s, 
chosen to be the same as that
for the time evolution in one dimension. In the one-dimensional
time evolution, the restriction on the time step due
to the explicit treatment of the nonlinear terms was
explored empirically, by varying $\Delta t$ and using a
value such that the algorithm was stable.  
We investigated
using a higher order (third order) Adams-Bashford scheme to verify
that the restriction on $\Delta t$ was limited by stability
as opposed to accuracy. 
The dynamically evolved
stationary state was compared with that obtained from
direct solution using a Newton-Raphson root-finding algorithm.
We determined the time step used in one dimension 
to be adequate for the time evolution in two dimensions as follows:
By using initial conditions uniform in the $x$-direction
(and equal to $5 \times 10^{-12}$ M for all species), thereby
reducing the two-dimensional time evolution to be effectively
one-dimensional, we verified that the asymptotic solution was the same as
that obtained in one dimension.
It is possible that implementing a numerical
algorithm adapted to solving stiff partial differential
equations (see Ref.\ \cite{ref:KLE_95} and references therein)
will improve the total execution time in the two-dimensional
evolution.
 

\section{Results}
\label{sec:result}
\subsection{Two-dimensional stationary solution}

In Fig.\ \ref{fig:fig1}, we
show density plots of the initial state, as well as 
the numerical solution for the
starch triiodide species after a total integration 
time of approximately $T=1000$ s,
with dark and light shading representing
low and high concentrations, respectively. 
In Figs.\ \ref{fig:fig2}(a)--(d), 
we show the time evolution of the $k^*$
mode and its higher order harmonics at 
the peak of the linear instability eigenvector 
($z=0.094$ cm) for the starch
triiodide species.  In Figs.\ \ref{fig:fig2}(e)--(h),
we plot the logarithm of the magnitude of these quantities.
(These plots are shown for the {\sl non-dimensionalized}
quantities.)
The slope of the linear segment in Fig.\ \ref{fig:fig2}(e),
for $t<\tau \sim 0.2$ is $5.129 \pm 0.012$ (corresponding to
$0.004616 \pm 0.000011$ s) agrees well
(to within one percent) with the linear growth rate, and
further verifies our linear stability results. 
In these plots, it is apparent that the saturated 
amplitudes of the higher order
harmonics are much smaller than that of $k^*$. 

To compare the size of the nonlinear perturbations in the
$x$-direction with the unperturbed profile in the $z$-direction,
we have plotted the $x$- and $z$-dependence of the
two-dimensional solution.
In Fig.\ \ref{fig:fig3}(a), we show 
the $x$-dependence at the peak of 
the linear instability eigenvector ($z=0.094$ cm), 
which approximates
a pure Fourier mode, verified by the
relatively small saturated amplitudes of the higher order
harmonics in Fig.\ \ref{fig:fig2}.
In Fig.\ \ref{fig:fig3}(b), 
the dashed and dotted lines denote the profiles in the
$z$-direction at $x=0.067 {\rm cm}=5\lambda$ and 
$x=0.073 {\rm cm}=5.5\lambda$, respectively.  
The solid line denotes the profile of the unperturbed
one-dimensional stationary state. We note that 
although the saturated amplitude of the transverse 
instability is comparable
to the variation of the one-dimensional stationary
state in the $z$-direction, its $x$-dependence is not
strongly nonlinear.

The results presented in this section can be summarized by three points.
First, we have presented the first numerical solution 
of two-dimensional patterns in the \textsl{boundary-fed} CDIMA
reaction-diffusion system using the LRE model.
Second, our numerical solution agrees qualitatively with patterns
observed in thin-strip reactors in experiments on the CIMA system,
for experimental conditions giving a single unstable front \cite{ref:BDD_95}.
The wavelength of the
solution presented here is $0.133$ mm, in agreement with
the $0.13-0.33$ mm experimentally observed range 
\cite{ref:OS_91}.
Finally, the numerical evolution in two dimensions confirms our
result from the linear stability analysis of the one-dimensional
stationary state along the gradients \cite{ref:ssmcc1}. 

\subsection{Bifurcation diagram}

\begin{figure}
   \centering \leavevmode
   \epsfxsize=6.5in
   \epsfbox{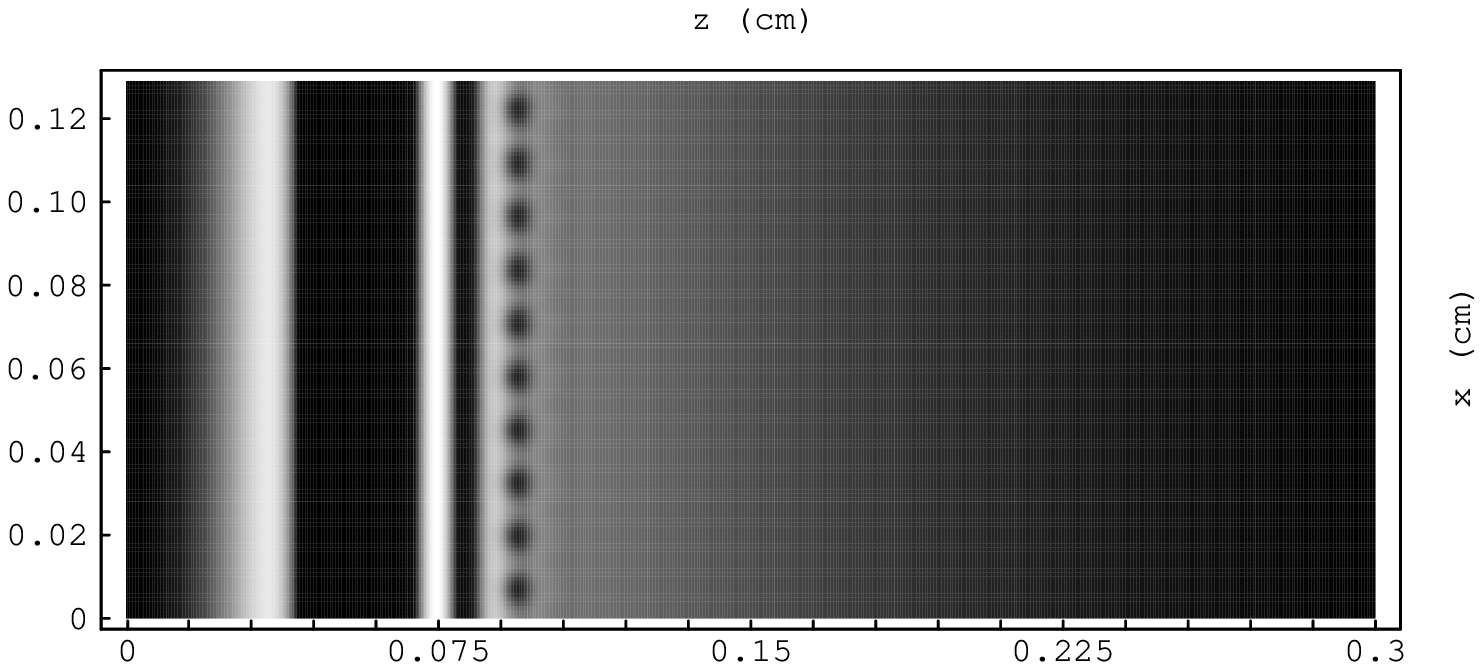}
   \vspace{0.2in}

   \centering \leavevmode
   \epsfxsize=6.5in
   \epsfbox{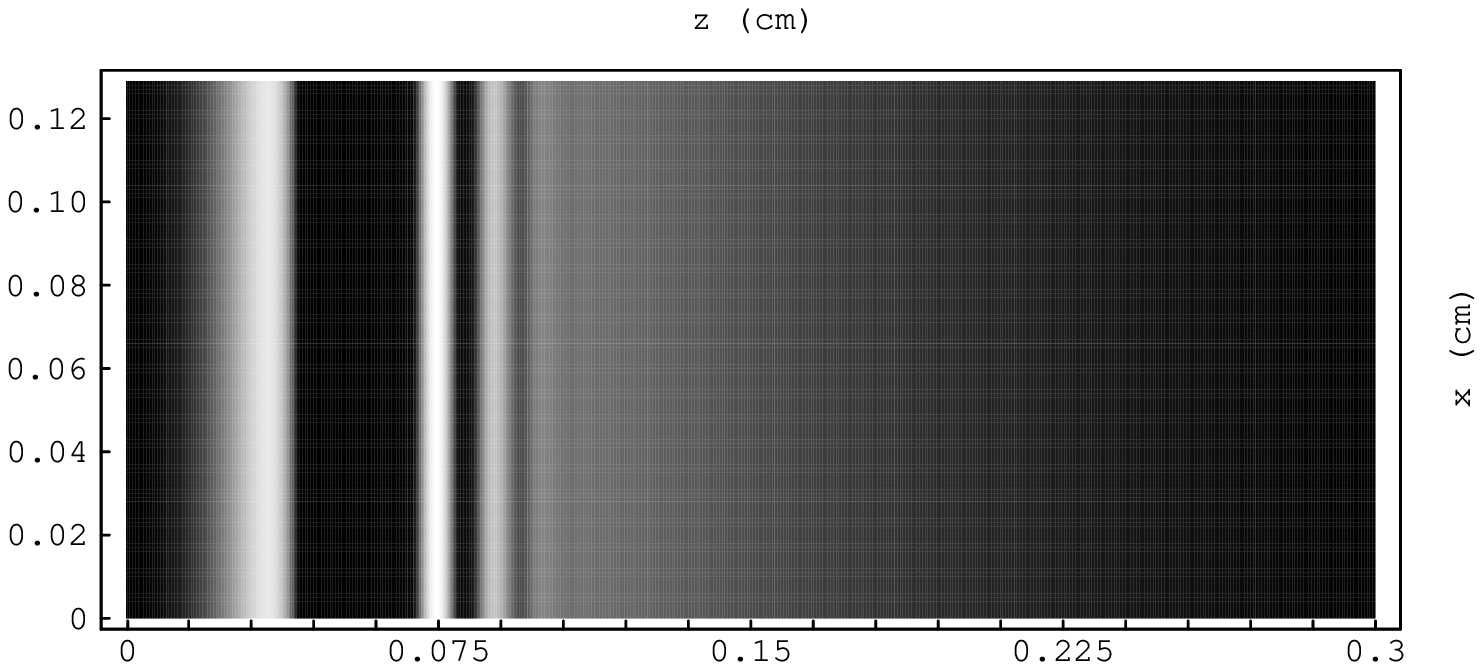}
   \vspace{0.2in}
   \caption[]{\small
Two-dimensional solution for starch triiodide with $\MA_{L}=0.0115$ M:  
The top figure
shows the numerical solution after $T=1000$ s of evolution time.
The bottom figure shows the initial condition.  Dark and light
shadings correspond to low and high concentrations, respectively.
}
   \label{fig:fig1}
\end{figure}

\begin{figure}
   \centering \leavevmode
   \epsfxsize=6.0in
   \epsfbox{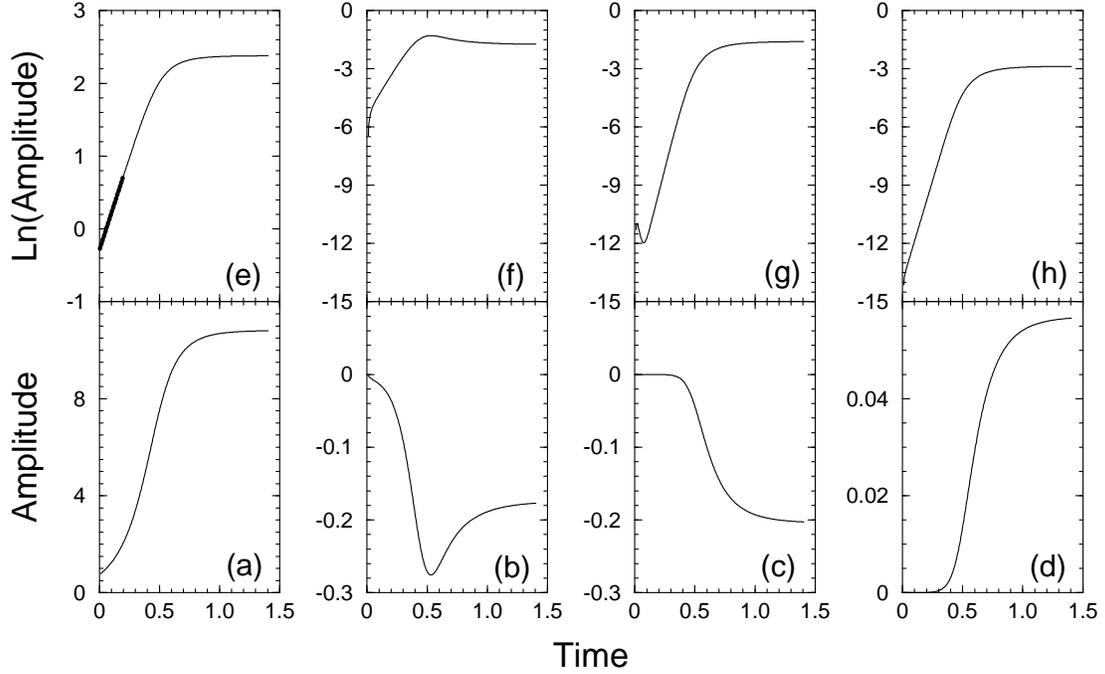}
   \vspace{0.2in}
   \caption[]{\small
Time-evolution of the most linearly unstable mode $k^*$ 
and its higher order harmonics:  
(a)--(d) 
show the values of the $k^*$, $2k^*$, $3k^*$ and $4k^*$ modes for the
starch triiodide species at the peak of the
most linearly unstable eigenvector ($z=0.094$ cm), as a function of 
time;  (e)--(h) show the logarithms of magnitudes 
of these amplitudes. The plots are shown
for non-dimensionalized quantities.   
(The time and concentration conversion factors 
are $9\times10^{-4}$ ${\rm s^{-1}}$ and $5\times10^{-5}$ M, respectively.)
The slope of the linear segment (heavy line) in plot (e)
for $t<0.2$ is $5.129 \pm 0.012$,
and agrees well with the growth rate $\lambda(k^*)=5.172$ 
from the linear stability analysis.
}
   \label{fig:fig2}
\end{figure}

\begin{figure}
   \centering \leavevmode
   \epsfysize=3.25in
   \epsfbox{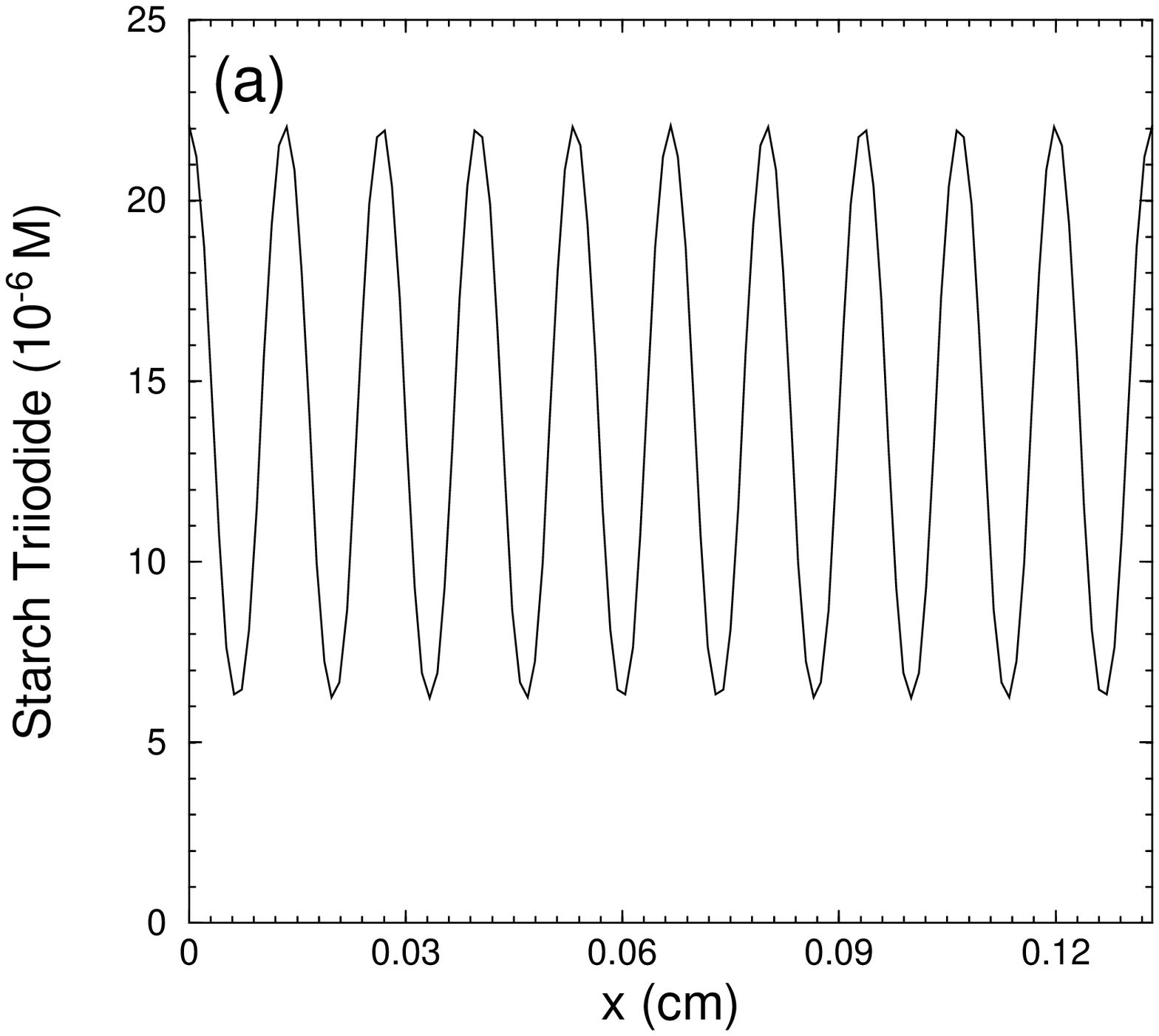}

   \vspace{0.1in}
   \centering \leavevmode
   \epsfysize=3.25in
   \epsfbox{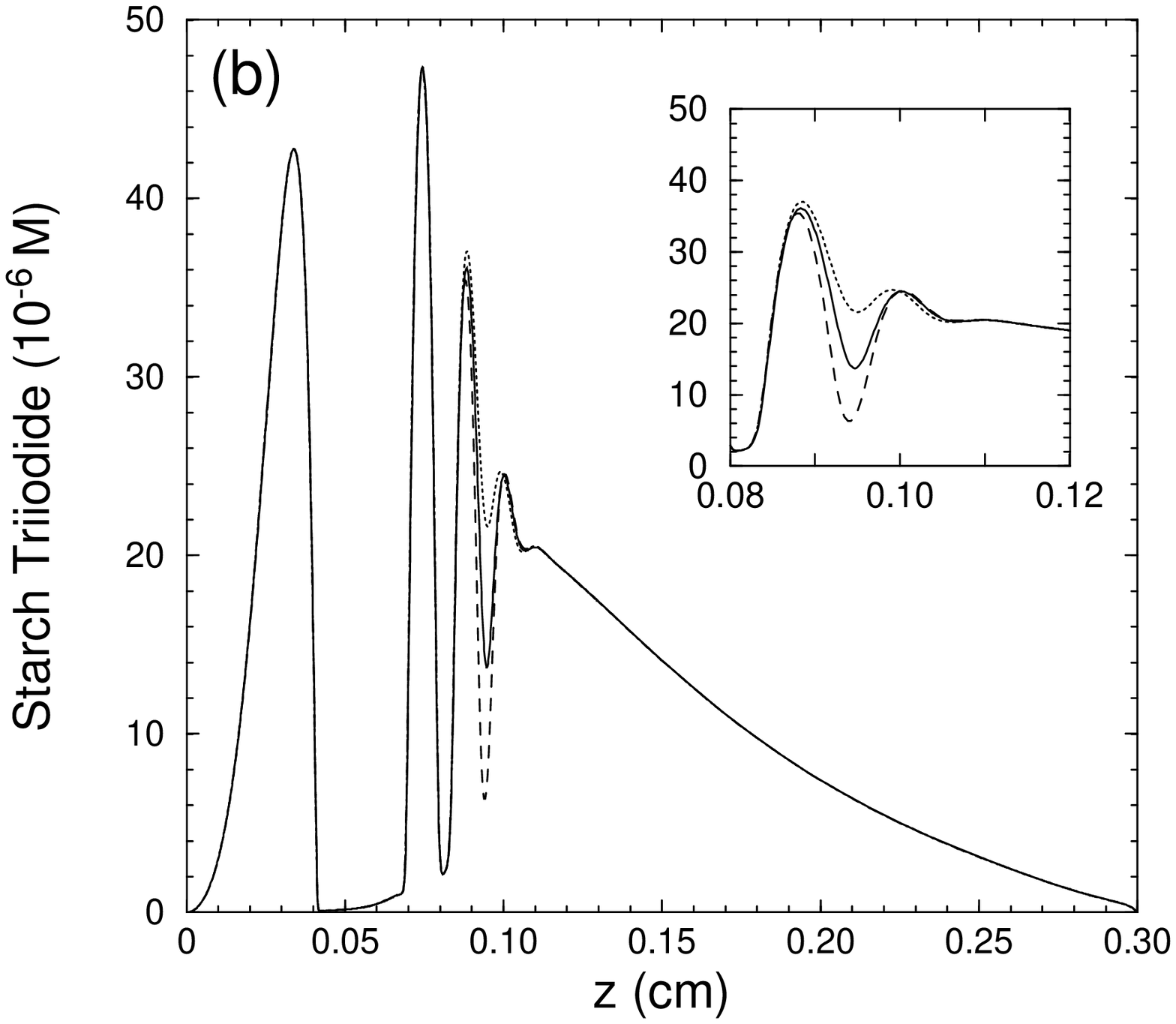}
   \caption[]{\small
Spatial dependence of the two-dimensional solution 
in the $x$- and $z$-directions for $\siiim$ species:  (a) The system
size in the $x$-direction is $0.133$ mm, constructed to
incorporate exactly ten wavelengths of the most
linearly unstable mode $k^*$. The profile in the
$x$-direction is plotted at $z=0.094$ cm. (b) The
dashed and dotted lines denote the profiles in the
$z$-direction at $x=0.067 {\rm cm}=5\lambda$ and 
$x=0.073 {\rm cm}=5.5\lambda$, respectively.  The solid
line denotes the profile of the unperturbed
one-dimensional stationary state. We note that the
saturated amplitude of the instability is comparable
to the variation in the profile of the one-dimensional stationary
state in the $z$-direction.
}
   \label{fig:fig3}
\end{figure}

The symmetry-breaking transition from Fig.\ \ref{fig:fig1}(a)
to Fig.\ \ref{fig:fig1}(b) is effectively one-dimensional, since
only a single layer is unstable over a range of values of
$\MA_{\small \rm L}$ control parameter, as we showed in
reference \cite{ref:ssmcc1}. In this earlier work,
we demonstrated the existence of (three) disjoint unstable ranges
as the $\MA_{\small \rm L}$ control parameter was continuously
varied from $4.0\times 10^{-3}$ M to $4.0 \times 10^{-2}$ M, 
consistent with experimental observations
in thin-strip reactors showing the appearance and subsequent
vanishing of a transverse instability as one of reservoir
concentrations was increased.  Here, we numerically investigate
the dependence of the saturated amplitude of the transverse instability on 
the $\MA_{\small \rm L}$ control parameter 
in the vicinity of the (lower) bifurcation point for one
of these unstable ranges 
($9.73 \times 10^{-3} {\rm M} < \MA_{\small \rm L} <1.25 \times 10^{-2} 
{\rm M}$). In the following, for convenience, we
report our results in \textsl{non-dimensionalized} 
units: the time conversion factor is 
$k_{1a}=9\times10^{-4}$ ${\rm s^{-1}}$,
and the concentration  conversion factor is 
$k_{1b}=5\times10^{-5}$ M.

First, the critical control parameter was determined numerically
from a linear fit to the maximum linear growth rate versus 
$\MA_{\small \rm L}$.  This value was found to be 
$\MA_{\small \rm c}=194.5226$.
Linear stability analysis of the stationary
state at this value of $\MA_{\small \rm L}$ yields a maximum growth rate of
$\lambda^* = 1.1054 \times 10^{-4}$.
This value of $\lambda^*$, which is expected to be zero, gives
a combined measure of the numerical uncertainties in the determination of
the stationary state at a particular value of $\MA_{\small \rm L}$ as well 
its linear stability. Hence, in principle, there would 
be error bars associated with values of $\MA_{\small \rm L}$,
equal to 
$\Delta \MA_{\small \rm L} = a \, \Delta \, \lambda = 3.0826 \times 10^{-4}$,
where $a$ is the slope of the linear fit from which $\MA_{\small \rm c}$
is extracted.
 
\begin{figure}
   \centering \leavevmode
   \epsfysize=4.0in
   \epsfbox{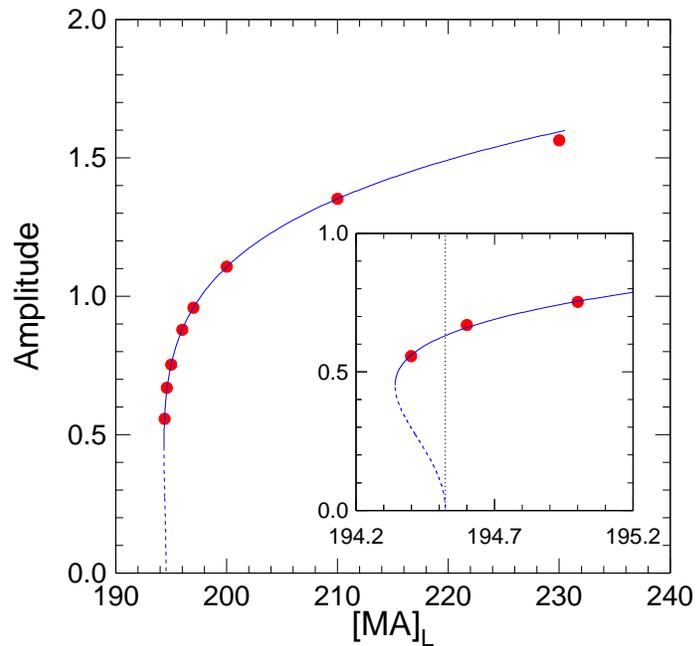}
   \caption[]{Bifurcation diagram:  The solid curve is the
computed fit; the broken curve corresponds to the unstable branch.
The inset shows
the vicinity of the saddle node bifurcation ($\MA_{\small \rm SN}=194.34$),
and the dotted vertical line denotes
linear threshold ($\MA_{\small \rm c}=194.5229$). }
   \label{fig:fig4}
\end{figure}

Fig.\ \ref{fig:fig4} shows the computed bifurcation diagram.
Starting in the supercritical regime, 
for each value of $\MA_{\small \rm L}$, as initial condition, we
use the corresponding one-dimensional stationary 
solution in the $z$-direction seeded with the most linearly unstable
eigenvector at approximately the 
saturated amplitude of the previous higher value
of control parameter (adiabatic approach).  The final converged
amplitudes were obtained from a least squares fit of the dynamical
evolution of the peak amplitude to an exponential plus a constant
offset, excluding initial points until the converged value did
not depend on the number of excluded points. 

Close to onset, the solution is given by:
\begin{equation}
u(x,z,t)=\overrightarrow{\delta u}_{k^*}(z) \cdot \left [
             A(t) \exp{(i k^* x)} + A^*(t) \exp{(-i k^* x)} \right],
\end{equation}
where $k^*$ is the most linearly unstable mode, and $A(t)$, $A^*(t)$
are complex conjugates.
The amplitude equation, which gives a universal description 
of the weakly nonlinear behavior 
and depends only on the symmetries of the problem 
(in this case, translational invariance in the $x$-direction) is
(at seventh order):
\begin{equation}
	\frac{\partial A}{\partial t} = \epsilon A 
                   + g_1 \left | A \right|^2 A 
                   + g_2 \left | A \right|^4 A 
                   + g_3 \left | A \right|^6 A,
\end{equation} 
where the coefficients depend on the specific system
under investigation, and 
$\epsilon \equiv \MA_{\small \rm L}-\MA_{\small \rm c}$
is the distance from onset.
Our results, described below, reveal a subcritical (first order) bifurcation.  

Fig.\ \ref{fig:fig4} shows a sixth order polynomial 
fit to the numerical data: 
$\MA_{\small \rm L} = \MA_{\small \rm c} - g_1 \left |A \right|^2 
                    - g_2 \left |A \right|^4 
                    - g_3 \left |A \right|^6$.
First, $\MA_{\small \rm c}$ was held fixed at
the linear threshold value, and $(g_1, g_2, g_3)$
were fitted for.  The goodness of the fit depended
on the number of points farthest from linear threshold
retained in the fit.  We determined that excluding
the last point $(\MA_{\small \rm L}=230.0)$ gave the best
fit.  The fitted parameters are:
\begin{eqnarray}
         g_1 & = & 1.7585,   \label{eq:g1fixed} \\
         g_2 & = & -4.0825,   \label{eq:g2fixed} \\
         g_3 & = & -0.82677.   \label{eq:g3fixed}
\end{eqnarray}
Since $g_1>0$, the instability does not saturate for
$\epsilon > 0$, and the bifurcation is subcritical.
We also investigated allowing all parameters, 
$(\MA_c, g_1, g_2, g_3)$, to float.
Again, excluding the last point produces
a fit with an offset $\MA_{\small \rm c}$, which is closest
to the linear threshold value. The
values of the fitted parameters in this case are:
$\left( \MA_{\small \rm c}, g_1, g_2, g_3 \right) = 
\left( 194.47, 1.5548, -3.8692, -0.89160 \right)$. 

\begin{figure}
   \centering \leavevmode
   \epsfysize=4.0in
   \epsfbox{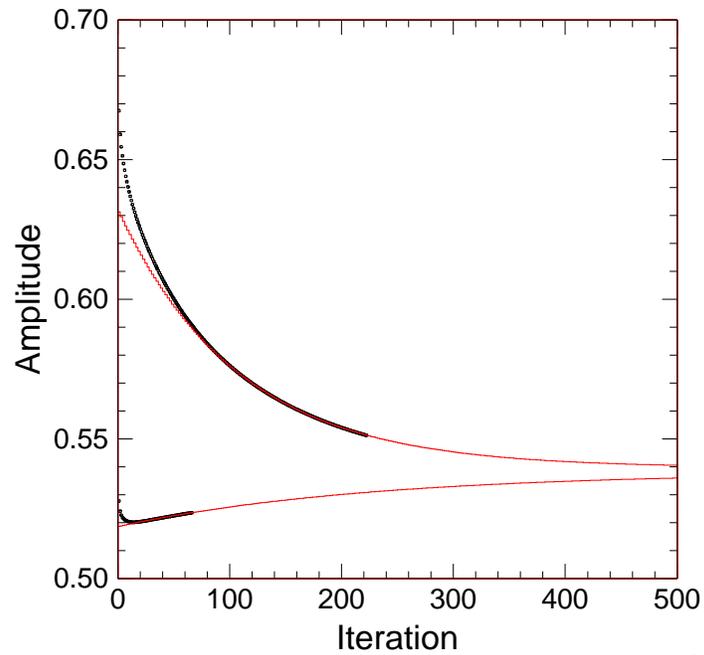}
   \caption[]{Convergence to finite amplitude below linear threshold, 
$\MA_{\small \rm L}=194.4$:  The closely-spaced circles denote the 
numerical time evolution, 
and the solid lines denote the computed fit to an exponential plus a constant
offset. Convergence from above and below to a finite amplitude is apparent.}
   \label{fig:fig5}
\end{figure}

Fig.\ \ref{fig:fig5} shows the convergence of
the peak amplitude of the fastest growing
linear mode for $\MA_{\small \rm L} = 194.4 $ in the subcritical region.
It shows convergence from above and below to a finite
amplitude instability.
The error for the converged amplitude is estimated
to be one-half of the difference between the
converged-from-above and converged-from-below values.
This error $(8.0 \times 10^{-4})$ is taken to be the same for all points,
eventhough the convergence from below was not repeated for all
points.\footnote{This convergence from above and below was also
confirmed for a point in the supercritical region, 
$\MA_{\small \rm L}=195.0$,
with even better agreement between the two fitted converged
values $(1.0 \times 10^{-4})$.}
We also confirmed the
decay of a \textsl{linear} perturbation at this same subcritical value.  
An exponential fit to the dynamical evolution
of the peak amplitude of the instability yields a
decay rate of $\lambda=-0.0416$, in good agreement with
the largest eigenvalue, $\lambda=-0.0429$.

The minimum value of $\MA_{\small \rm L}$ below which a finite amplitude
instability does not exist, corresponding to the saddle-node
bifurcation, can be computed from the fitted
parameters.  Using the parameter values given in
Eq.\ \ref{eq:g1fixed}--\ref{eq:g3fixed},  $\MA_{\small \rm SN}$ is
found to be:
\begin{equation}
        \MA_{\small \rm SN} = 194.34.
\end{equation}
The inset in Fig.\ \ref{fig:fig4} shows this turning point. 
For $\MA_{\small \rm L}=193.0$ below this value, we explicitly
verified decay to zero of an initial perturbation
with amplitude $A=0.7756$.

We note that the transition is ``weakly'' subcritical.  This is
characterized by the small range of control parameter 
below linear threshold, approximately equal to $0.18$ 
($9 \times 10^{-6}$ M), for which
a finite amplitude instability exists, in comparison with
the linearly unstable range, $55.4$ ($2.77 \times 10^{-3}$ M),
determined in our earlier work \cite{ref:ssmcc1}.
The weakly subcritical nature of the transition implies that
a linear stability analysis
of the one-dimensional structures along the gradients
\cite{ref:ssmcc1} does have utility in predicting
the existence of a transverse instability
for given reaction parameters and boundary conditions over
a wide range of control parameters (in the supercritical
regime).


\section{Conclusion and Discussion}
\label{sec:conclude}

To summarize, we have carried out a two-dimensional numerical simulation 
of the boundary-fed CDIMA reaction-diffusion system based on the 
realistic LRE model for this system.  Our results are qualitatively
similar to those seen in experiments on the CIMA reaction-diffusion
system, and support the earlier work \cite{ref:ssmcc1} in which we studied the 
linear instability of the boundary-fed CDIMA system to transverse Turing patterns.

Numerical studies by Jensen
{\sl et al.} \cite{ref:JPDB_93,ref:JPMDB_93,ref:JMBD_96} 
on the two-variable LRE model 
with {\sl uniform} backgrounds  
have found the transition to stripes in one
and two dimensions to be subcritical.
Our results demonstrate this transition
to also be subcritical in the \textsl{boundary-fed} LRE model.
This prediction
and computed bifurcation diagram can be directly verified
by experiments based on the CDIMA system. 

The subcritical nature of the transition to stripes
makes the LRE model qualitatively different from other
abstract reaction-diffusion models hitherto used to 
study Turing patterns.  
For example, in the ramped Brusselator,
the transition to stripes has been shown to be 
supercritical \cite{ref:B_88}.  
Jensen \textsl{et al.}
have investigated the propagation of fronts separating
the homogeneous steady state from the Turing structure in one and two
dimensions using the uniform LRE model.  The subcriticality
allows for the existence of a range
values of control parameter for which the front
velocity vanishes, allowing an infinite number of 
stable steady inhomogeneous structures. 
Despite the weakly subcritical nature of the transition, 
it would be interesting to similarly investigate front propagation and
formation of localized (quasi one-dimensional) states in the 
boundary-fed system.

In experimental geometries (disc reactors) 
where the dimensions of the reactor
transverse to the gradients are large, 
the analog of the one-dimensional row of spots
which develops in our numerical simulation 
and in experiments using thin-strip reactors
is a two-dimensional ``monolayer''.
Dufiet {\sl et al.} \cite{ref:DB_96} have pointed out that these
monolayers, which are confined by a strong transverse gradient of reservoir
chemical concentrations,
must be distinguished from {\sl genuine} two-dimensional
structures with uniform control parameters.
Pattern selection in genuine 
two- and three-dimensional
systems has been studied analytically and numerically
using abstract reaction-diffusion models \cite{ref:DDBW_92}. 
However, it is not practical to generate sustained
genuine structures experimentally. 
In the context
of a model reaction-diffusion system, Dufiet \textsl{et al.} have shown
that in genuine two-dimensional systems and monolayers,
the stripe-hexagon competition is similar close to onset.
They find, however, that far from onset,
hexagonal phases in monolayers are restabilized
due to their interaction with a longitudinal ($k=0$) instability.
The latter finding
is consistent with earlier theoretical predictions
\cite{ref:Price_94,ref:Price_95}, as well as experiments in bevelled
disc reactors \cite{ref:DDRK_96}.

It would
be interesting to numerically investigate pattern selection for
monolayers in the LRE model of the CDIMA system in the range
of boundary conditions and reaction parameters accessible to
experiments, allowing in principle direct comparison with experimental results.
This would require extension of our numerical computation to
three dimensions.

%
\acknowledgments
This work was supported by the National Science Foundation under 
Grant No. DMR 9311444, and by a generous award of computer time from
the Center for Advanced Computing Research at Caltech. 
We thank Ruben Krasnopolsky for invaluable advice with the parallelization.


\pagebreak
\begin{table}[hbt]
  \begin{center}
  \caption[Kinetic constants used in the LRE model]{
             Kinetic constants for the CDIMA system.
            }
  \label{table:T1}
  \begin{tabular}{ccc}
   Rate or diffusion constant & Dimensions  & Value \\
  \hline
    & & \\
    $k_{1a}$  & (s$^{-1}$)          & $9  \times 10^{-4}$ \footnotemark[1]   \\
    $k_{1b}$  & (M)                 & $5  \times 10^{-5}$ \footnotemark[1] \\
    $k_{2}$   & (M$^{-1}$s$^{-1}$)  & $1  \times 10^3$ \footnotemark[1]    \\
    $k_{3a}$  & (M$^{-2}$s$^{-1}$)  & $1.2\times 10^2$ \footnotemark[1]    \\
    $k_{3b}$  & (s$^{-1}$)          & $1.5\times 10^{-4}$ \footnotemark[1] \\
    $h$       & (M$^2$)             & $1.0\times 10^{-14}$ \footnotemark[1] \\
    $k_+$     & (M$^{-2}$s$^{-1}$)    & $6.0\times 10^5$ \footnotemark[2]    \\
    $k_-$  & (s$^{-1}$)    & $1.0$ \footnotemark[2]    \\
    $D_{\im}$ & (cm$^2$s$^{-1}$)      & $7.0\times 10^{-6}$ \footnotemark[3] \\
    $D_{\clom}$ & (cm$^2$s$^{-1}$)    & $7.0\times 10^{-6}$ \footnotemark[3] \\
    $D_{\ii}$ & (cm$^2$s$^{-1}$)      & $6.0\times 10^{-6}$ \footnotemark[1] \\
    $D_{\ma}$ & (cm$^2$s$^{-1}$)      & $4.0\times 10^{-6}$ \footnotemark[1] \\
    $D_{\clo}$ & (cm$^2$s$^{-1}$)     & $7.5\times 10^{-6}$ \footnotemark[1] \\
    $D_{\hp}$ & (cm$^2$s$^{-1}$)     & $1.0\times 10^{-5}$ \\
    $K[\rm S]_o$ & (M$^{-1}$)      & $6.25\times 10^{4}$ \footnotemark[4]\\
  \hline
\multicolumn{3}{l}{
\footnotemark[1]{From \cite{ref:LE_91} at $7^\circ {\rm C}$}; 
\footnotemark[2]{From \cite{ref:KLE_95} at $4^\circ {\rm C}$}; 
\footnotemark[3]{From \cite{ref:LKE_92} at $4^\circ {\rm C}$}; 
\footnotemark[4]{From \cite{ref:LE_95} at $4^\circ {\rm C}$}.
}
  \end{tabular}
  \end{center}
\end{table}


\end{document}